\documentclass{elsart}

 \usepackage{graphicx}
\newcommand{\km}{{k_*}} 

\begin{document}

\begin{frontmatter}

% Title, authors and addresses

% use the thanksref command within \title, \author or \address for footnotes;
% use the corauthref command within \author for corresponding author footnotes;
% use the ead command for the email address,
% and the form \ead[url] for the home page:
% \title{Title\thanksref{label1}}
% \thanks[label1]{}
% \author{Name\corauthref{cor1}\thanksref{label2}}
% \ead{email address}
% \ead[url]{home page}
% \thanks[label2]{}
% \corauth[cor1]{}
% \address{Address\thanksref{label3}}
% \thanks[label3]{}

\title{Excitation of travelling multibreathers in anharmonic chains}

% use optional labels to link authors explicitly to addresses:
% \author[label1,label2]{}
% \address[label1]{}
% \address[label2]{}

\author[l0]{Ramaz Khomeriki}
%\ead{khomeriki@hotmail.com}
\author[l1,l2]{Stefano Lepri}
%\ead{stefano.lepri@unifi.it}
\author[l1,l2,l3]{Stefano Ruffo}
%\ead{ruffo@avanzi.de.unifi.it}

\address[l0]{Department of Physics, Tbilisi State University, \\
Chavchavadze ave. 3, Tbilisi 380028, Georgia}
\address[l1]{Dipartimento di Energetica ``Sergio Stecco" \\
Universit\'a di Firenze, Via S. Marta, 3 - I-50139 Firenze, Italy}
\address[l2]{Istituto Nazionale di Fisica della Materia, Unit\`a di Firenze}
\address[l3]{Istituto Nazionale di Fisica Nucleare, Sezione di Firenze}

\begin{abstract}
We study the dynamics of the ``externally" forced and damped Fermi-Pasta-Ulam
(FPU) 1D lattice. The forcing has the spatial symmetry of the Fourier mode
with wavenumber $p$ and oscillates sinusoidally in time with the frequency $\omega$. 
When $\omega$ is in the phonon band, the $p$-mode becomes modulationally 
unstable above a critical forcing, which we determine 
analytically in terms of the parameters of the system.
For $\omega$ above the phonon band, the instability of the $p$-mode leads
to the formation of a {\it travelling multibreather}, that, in the
low-amplitude limit could be described in terms of soliton solutions of a suitable
driven-damped nonlinear Schr\"odinger (NLS) equation. 
Similar mechanisms of instability could show up in easy-axis 
magnetic structures, that are governed by such NLS equations. 
\end{abstract}

\end{frontmatter}

% main text
\section{Introduction}
\label{se:Introduction}

Chains of anharmonic oscillators constitute an
ideal testing ground for the different theoretical approaches
to the study of the dynamics of many-degrees-of-freedom systems.
For instance, the first numerical study of the {\it approach to
equilibrium} was performed by Fermi, Pasta and Ulam (FPU), initializing the chain
in a long wave-length state and looking at mode energy sharing at 
long-times~\cite{fermi}. Later on, the role of certain exact nonlinear standing
and travelling wave solutions of such Hamiltonian lattices
was discussed in relation with chain excitations at shorter 
wavelengths~\cite{kosevich,poggi,perez}. This led to the discovery of
{\it chaotic breathers}~\cite{ruffo1}, spatially highly localized and erratically moving
objects, which appear in the process of thermalization
at small energy densities (these were in fact previously found in Ref.~\cite{burla}
without emphasizing too much their role for the approach to equilibrium).
This discovery motivated a series of papers~\cite{lepri,corso,mirnov} where the
formation, lifetimes and destruction of these ``intrinsically localized" structures
were studied in full detail.
Very recently, a different approach was followed in which, instead of looking at
the fate of a specific initial state of such FPU lattices, e.g. the highest frequency zone-boundary
Fourier $\pi$-mode, a driving term with this spatial symmetry was directly added to the
Hamiltonian equations of motion~\cite{ssr}. Damping was included
to allow the formation of stationary states and it was taken to be weak 
to remain close to the Hamiltonian case. 
As a result the zone-boundary mode was found to be stable and locked to the 
driving field below a certain critical forcing, which is indeed analytically calculable. 
Above such forcing, a {\it standing modulated wave} forms for driving frequencies
that are below the band edge, while a {\it static multibreather} (a train of breathers)
develops for forcing frequencies above the band edge.
The patterns emerging from the instability of the zone-boundary mode are 
always almost static, because the group velocity of the carrier wave is zero. 
They can acquire a nonzero velocity due to other nonlinear 
mechanisms~\cite{lepri}.

In this paper we study a damped FPU 1D lattice forced by a travelling wave field with
wavenumber $p\neq\pi$.
An instability mechanism similar to the one mentioned above leads to the formation of 
{\it travelling modulated waves} for in-band frequencies.
When the forcing frequency is instead  above the phonon
band, a {\it moving multibreather} arises after instability
of the $p$-mode. This state can indeed be thought as a moving
envelope soliton train (the carrier wave having the symmetry of
the $p$-mode), since the envelope of these ``intrinsically localized"
moving objects can be put into a close relation with the 
soliton solutions of a suitable externally driven and damped
nonlinear Schr\"odinger (NLS) equation.
Such type of equation arises in the small amplitude limit
of the ac driven damped sine-Gordon (SG) system; it is discussed 
in the literature and its analytical and numerical solutions are 
known~\cite{baras,baras2,mazor}. 

In Section~\ref{Sec2} we introduce a Fourier mode representation
of the FPU lattice and we study it in the {\it rotating wave}
approximation and for weak damping, obtaining an explicit analytical
expression for the critical amplitude $f_{cr}$ in terms of the parameters 
of the system. Section~\ref{Sec3} is devoted to the description
of the travelling multibreather, both numerically and, analytically, in terms   
of a suitable NLS equation. In Section \ref{Sec4} we draw some conclusions.

\section{Stability of the travelling wave}
\label{Sec2}

The equations of motion of the ``externally driven" damped FPU chain 
read as follows:
\begin{eqnarray}
\ddot u_n=u_{n+1}+u_{n-1}-2u_n +(u_{n+1}-u_n)^3
 && +(u_{n-1}-u_n)^3 \nonumber \\  && -\gamma
\dot u_n+f\cos (\omega t+p n),
\label{fpu}
\end{eqnarray}
where $u_{n}$ is the displacement of $n$-th oscillator with respect to its
equilibrium position. Periodic boundary conditions,  $u_{n+N}=u_n$, are
assumed, with $N$ being the number of oscillators.  Dimensionless units are
used such that the masses, the linear and nonlinear force constants and the 
lattice spacing are taken equal to unity. The forcing and damping strengths 
are gauged by the parameters $f$ and $\gamma$, respectively; $\omega$ and $p$ are
the driving frequency and wavenumber. For technical reasons, to be
discussed below, we restrict to $\pi/2 <|p| <\pi$. As in Ref.~\cite{ssr}
we use the following expansion:
\begin{equation}
u_n=\frac{1}{2}\sum\limits_k\left[a_k \, e^{i(\omega t+kn)} +
a_{-k}^+ \,e^{-i(\omega t-kn)}
\right],
\label{aq}
\end{equation}
where $a_k$ are complex mode amplitudes and $-\pi < k \leq \pi$
the corresponding wavenumbers.
The equations of motion for the amplitudes $a_k$'s are obtained by substituting
Eq.~(\ref{aq}) into Eq.~(\ref{fpu}). Similarly to what is done for
the undamped case~\cite{flach}, a considerable simplification is achieved
by neglecting higher-order harmonics that are produced by the cubic force terms.
This is the so-called {\it rotating wave approximation}, that is valid for
small amplitudes, $|a_k|\ll 1$. Moreover, in the limit of weak damping 
$\gamma \ll \omega$ and neglecting inertial terms $\ddot{a}_k$,
we find the following set of approximate equations:
\begin{equation}
-2i\omega \dot a_k -i\omega\gamma
a_k=(\omega_k^2-\omega^2)a_k+f\delta_{k,p}
+6\sum\limits_{q_1,q_2}G_{q_1,q_2}^k a_{q_1}a_{q_2}a_{q_1+q_2-k}^+,
\label{eqa}
\end{equation}
where
\begin{equation}
\omega_k^2=2(1-\cos k), 
\end{equation}
$\delta_{k,p}$ is equal to one for $k=p$ and zero otherwise and
\begin{eqnarray}
G_{q_1,q_2}^k=&&\frac{1}{4} [1+\cos (q_1+q_2)+\cos (k-q_2)
+\cos (k-q_1) \nonumber \\
&&-\cos k-\cos q_1 -\cos q_2-\cos (k-q_1-q_2) ].
\label{Gqqk}
\end{eqnarray}
Let us begin by considering solutions where only the $p$-mode is excited. 
Such solutions, corresponding to nonlinear travelling waves with velocity
$v=\omega/p$, are numerically observed to exist and to be stable below a 
certain critical forcing $f_{cr}$. Indeed, when we initialize the lattice 
at equilibrium and we impose random velocities, all modes with $k \ne p$ 
damp out on a time-scale set by the value of $\gamma^{-1}$ while 
$|a_p|$ rapidly grows and finally approaches a constant value. 
In this asymptotic regime the system is well described by the
``internal dynamics" of the $p$-mode, given by the following equation:
\begin{equation}
-2i\omega\dot{a}_p-i\omega\gamma a_p=
(\omega_p^2-\omega^2)a_p+f+\frac{3}{4}\omega_p^4a_p|a_p|^2~.
\label{internal}
\end{equation}
The asymptotic amplitude $a_{p}$ is then obtained by solving for 
the stationary solution of Eq.~(\ref{internal}), which originates the
algebraic equation below:
\begin{equation}
a_{p}=\frac{f}{\omega^2-\omega_p^2-\frac{3}{4}\omega_p^4|a_p|^2
-i\gamma\omega},
\label{fixedpoint}
\end{equation}
which can be numerically solved for both the modulus and the phase of
$a_p$. In particular the squared modulus of $a_p$, $z=|a_p|^2$, is the 
solution of the cubic equation: 
\begin{equation}
\frac{9}{16}\omega_p^8 z^3 -\frac{3}{2}\omega_p^4(\omega^2-\omega_p^2) z^2 +
\left[(\omega^2-\omega_p^2)^2+\gamma^2\omega^2\right] z = f^2 \quad.
\label{cub}
\end{equation}
In reality, also the higher odd harmonics of $p$ will be excited, but they
can be neglected when $|a_p| \ll 1$. 
One can easily ascertain, as it also happens for the $p=\pi$ case~\cite{ssr}, 
that Eq.~(\ref{cub}) admits a single real root only for $|\omega|<\omega_*$ where 
\begin{equation}
\omega_*\simeq\omega_p+\sqrt{3}\gamma/2, \label{crit}
\end{equation}
while three distinct roots may otherwise exist. We will not analyse in this
short note what happens when three solutions of the consistency 
equation (\ref{fixedpoint}) coexist for a given parameter set. 
It turns out that, the destabilization of the $p$-mode can be described 
by considering the lowest amplitude solution of~(\ref{internal}).

The $p$-mode nonlinear solution is stable for $f<f_{cr}(\omega,\gamma,p)$.
It is possible to get an analytic expression for $f_{cr}(\omega,\gamma,p)$
in the range $0<\omega<\omega_*$.  
This is accomplished by solving the following set of equations, that one
obtains when linearizing Eqs.~(\ref{eqa}) around the stationary 
$p$-mode solutions (i.e. neglecting all mode interaction terms which are 
nonlinear in the $a_k$'s)
\begin{eqnarray}
&&-2i\omega \dot a_k-i\omega\gamma a_k=(\tilde\omega_k^2-\omega^2)a_k
+\frac{3}{2}\omega_p^2B_{k,p} a_p^2 a_{2p-k}^+ \nonumber \\ 
&&-2i\omega \dot a_{2p-k}^+-i\omega\gamma a_{2p-k}^+=-(\tilde\omega_{2p-k}^2-\omega^2)a_{2p-k}^+
-\frac{3}{2}\omega_p^2B_{k,p} (a_p^+)^2 a_{k}, 
\label{linearized}
\end{eqnarray}
where
\begin{equation}
\tilde\omega_k^2 =(1+\frac{3}{2}\omega_p^2|a_p|^2) \omega_k^2, \qquad
B_{k,p}=\cos (k-p)- \cos p
\label{coefs}
\end{equation}
are the frequency of the $k$-th mode shifted by the interaction
with the $p$-mode and $B_{k,p}$ derive from $G_{q_1,q_2}^k$ 
in Eq.~(\ref{Gqqk}), when the interaction is restricted to the
mode subset $k,p,2p-k$.
Since $B_{k,p}=B_{2p-k,p}$, the two Eqs.~(\ref{linearized}) can be
obtained one from the other for $k \to (2p-k)$, hence they can be 
solved looking for symmetric solutions of the form 
$a_k \sim a_{2p-k}\sim\exp(\nu_k t)$.
The relevant branch of the eigenvalue spectrum reads
\begin{eqnarray}
\nu_k = &&
-{\gamma \over 2} + i\frac{\tilde\omega_k^2-\tilde\omega_{2p-k}^2}{4\omega}\nonumber \\
&&+\frac{1}{2|\omega|}\sqrt{\frac{9}{4}\omega_p^4B_{k,p}^2|a_p|^4 -
\left(\omega^2-\frac{1}{2}(\tilde\omega_k^2+\tilde\omega_{2p-k}^2)\right)^2}~.
\label{nuk}
\end{eqnarray}
The growth rate $Re\{\nu_k\}$ is maximal when the square root in the above 
expression attains its maximum value, i.e. 
when the ``resonance" condition 
\begin{equation}
2\omega^2=\tilde\omega_k^2+\tilde\omega_{2p-k}^2 \label{rez}
\end{equation}
holds. The latter, together with the expression for $\tilde\omega_k^2$ in~(\ref{coefs}), 
allows to get the following equation for the wavenumber $\km$ of the most unstable mode
\begin{equation}
\cos (p-\km)=\frac{1}{\cos p}\left(1-\frac{\omega^2}{2+3\omega_p^2|a_p|^2)}\right)~,
\label{km}
\end{equation}
in terms of the amplitude of the $p$-mode.
Moreover, if we let $Re\{\nu_k\}=0$ in Eq.~(\ref{nuk}) and use the definition
of $B_{k,p}$ in (\ref{coefs}), we obtain the following equation for the
squared amplitude of the $p$-mode:
\begin{equation}
|a_p|^2=\frac{2\gamma\omega}{3\omega_p^2|\cos(\km-p)-\cos p|}~.
\label{ppp}
\end{equation}
Solving the set of  Eqs.~(\ref{km}) and~(\ref{ppp}) for both $\km$ and $|a_p|$ one
obtains the critical amplitude $|a_p|_{cr}$ of the $p$-mode, above which modulational 
instability occurs. Finally, replacing the expression
for $z_{cr}=|a_p|_{cr}^2$ in Eq.~(\ref{cub}), we obtain $f_{cr}(\gamma,\omega,p)$, as announced
above. We do not display its explicit expression because of the rather 
bulky form. All this is of course valid when only one solution of the cubic
equation~(\ref{cub}) is present, i.e. for $\omega<\omega_*$. 
Moreover, Eq.~(\ref{km}) has a solution in this full $\omega$ range only when
$\pi/2 < |p| < \pi$, as can be easily checked. In this range
of parameters we can get rather general expressions both for the stable $p$-mode
pattern and for the expression of the critical forcing $f_{cr}$.

When $f>f_{cr}$ the $p$-mode is modulationally unstable and, presumably, as for
the $p=\pi$ case~\cite{ssr}, the system saturates asymptotically, due
to nonlinear effects, into a state where the triplet of modes $\km,p,2p-\km$ 
is excited, resulting into a more complex {\it travelling modulated wave}. Its 
expression can possibly be computed, solving for the stationary solution of 
the coupled equations for this triplet of modes. This calculation is out
of the scope of the present short note.

\section{Travelling multibreathers}
\label{Sec3}

\begin{figure}
%\vspace {6 truecm}
\centering{\includegraphics[width=0.65\textwidth,height=0.45\textheight,angle=-90]{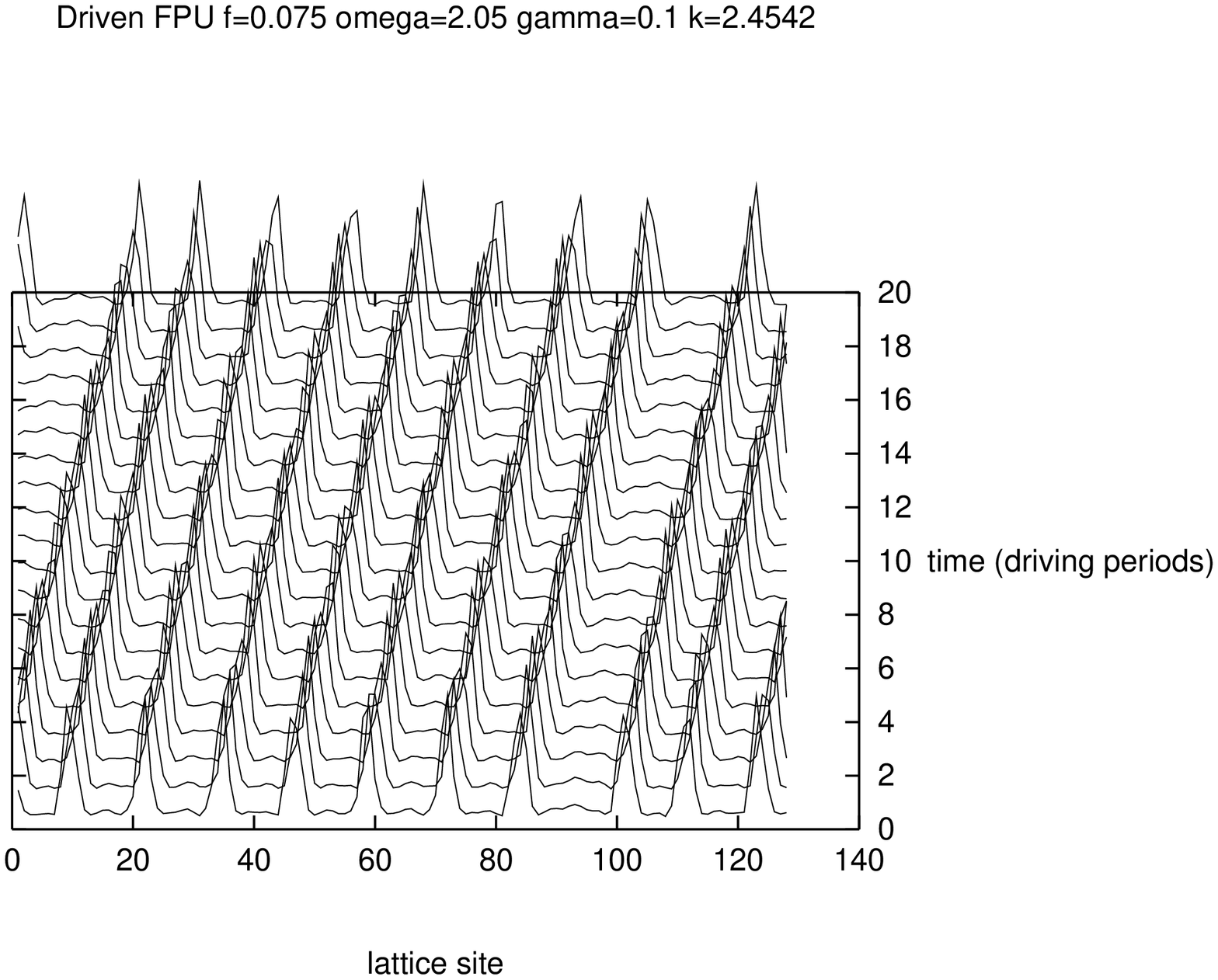}}
\caption{Travelling multibreather pattern generated after the modulational instability 
of the $p$-mode for a lattice of $N=512$ sites. Here, $\omega=2.05$, 
$f=0.075$, $p=2.4542$.}
\label{multibre}
%\bigskip
\end{figure}

\begin{figure}
%\vspace {6 truecm}
\centering{\includegraphics[width=0.65\textwidth,height=0.45\textheight,angle=-90]{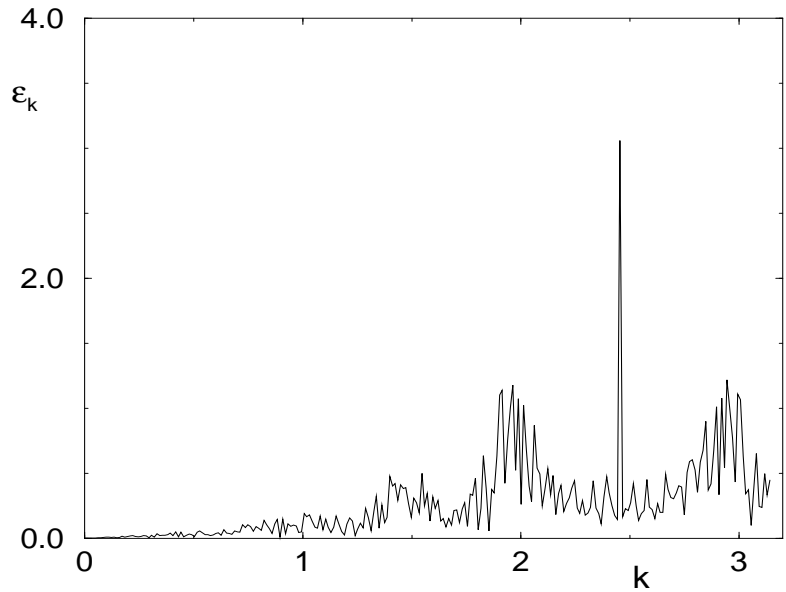}} 
\caption{Spatial spectrum of the travelling multibreather. 
$\epsilon_k=|\dot{U}_k|+\omega^2|U_k|$, where $U_k$ is the $k$-th component of the 
Fourier spectrum of the displacement field $u_n$. Same parameters as 
in Fig.~\ref{multibre}.}
\label{spectrum}
%\bigskip
\end{figure}

As it was mentioned above, in the range $|\omega|>\omega_*=\omega_p+\sqrt{3}\gamma/2$,
three roots of Eq.~(\ref{cub}) are present and the instability process is
determined by the ``internal" dynamics of the $p$-mode, as defined by 
Eq.~(\ref{internal}). After instability, several modes grow at the
same time and no periodic pattern develops. However, numerical simulations
show that a {\it travelling multibreather} can 
form just above threshold. An example is shown in Fig.~\ref{multibre}, where
we plot the local energy
\begin{eqnarray}
h_n = \frac{1}{2}\dot u_n^2 +
&&\frac{1}{2}\left[(u_{n+1}-u_{n})^2+(u_{n}-u_{n-1})^2 \right]+
\nonumber \\
&&\frac{1}{4}\left[(u_{n+1}-u_{n})^4+(u_{n}-u_{n-1})^4 \right]~ ,
\end{eqnarray}
vs. the lattice position sampled at the period of the forcing. 
The corresponding spatial Fourier spectrum is shown in Fig.~\ref{spectrum}.
The broad band structure of the spectrum reflect the non perfect periodic
arrangement of the localized peaks in Fig.~\ref{multibre}.

Such states can be described in terms of soliton solutions of
an associated suitable driven-damped nonlinear Schr\"odinger (NLS) equation. 
Let us first make the following definition:
\begin{equation}
u_n=\frac{1}{2}\left[a_p(n,t)e^{i(\omega t+pn)} +
a_{p}^+(n,t)e^{-i(\omega t+pn)}\right],
\label{aq1}
\end{equation}
where $a_p$ and its conjugate are smooth functions of $n$. Such an assumption is possible
if the wavepacket is concentrated around the driving mode $\Delta k\ll\pi$. Substituting
(\ref{aq1}) into the equations of motion ~(\ref{fpu}) one gets:
\begin{eqnarray}
(\omega_p^2-\omega^2)a_p && +2i\omega\left(\frac{\partial a_p}{\partial t}-
v\frac{\partial a_p}{\partial n}\right) \nonumber \\
&&+\frac{\omega_p^2}{4}\frac{\partial^2 a_p}{\partial n^2}+\frac{3}{4}\omega_p^4
a_p|a_p|^2=-i\omega\gamma a_p-f~. \label{ac}
\end{eqnarray}
After performing the following re-scalings:
\begin{eqnarray}
t'=\frac{\omega^2-\omega_p^2}{2\omega}t, \qquad \xi=\frac{2\sqrt{\omega^2-\omega_p^2}}{\omega_p}
(n-vt),\nonumber \\
\Psi=\sqrt{\frac{3}{8}}\frac{\omega_p^2}{\sqrt{\omega^2-\omega_p^2}}e^{it'}a_p(\xi,t'),
\qquad \gamma'=\frac{\omega}{\omega^2-\omega_p^2}\gamma, \nonumber \\
h=\sqrt{\frac{3}{8}}\frac{\omega_p^2}{(\omega^2-\omega_p^2)^{3/2}}f~,\nonumber
\end{eqnarray}
and choosing a reference frame moving with 
velocity $v=\partial \omega_p/\partial p=\sin p/\omega_p$, Eq.(\ref{ac})
reduces to the  well studied ``externally" driven (or ac driven) damped 
NLS equation~\cite{baras,baras2}:
\begin{equation}
i\frac{\partial\Psi}{\partial t'}+\frac{\partial^2\Psi}{\partial\xi^2}+2\Psi|\Psi|^2
=-i\gamma'\Psi-he^{it'}~. \label{nls}
\end{equation}
Exact soliton solutions of this equation can be obtained for $\gamma'=0$,
see Eqs.(37-40) of Ref.~\cite{baras}. Moreover, multisoliton solutions
are also derived in Ref.~\cite{baras2}.
What we observe in Fig.~\ref{multibre} might well be a superposition
of such solutions to form a train of ``intrinsically localized"
structures. However, one should bear in mind that NLS solutions can describe
only low amplitude states. Therefore, they can be only a fair approximation
of the pattern displayed in Fig.~\ref{multibre}, which shows high 
amplitude localized peaks.

\begin{figure}
%\vspace {6 truecm}
\centering{\includegraphics[width=0.65\textwidth,height=0.45\textheight,angle=-90]{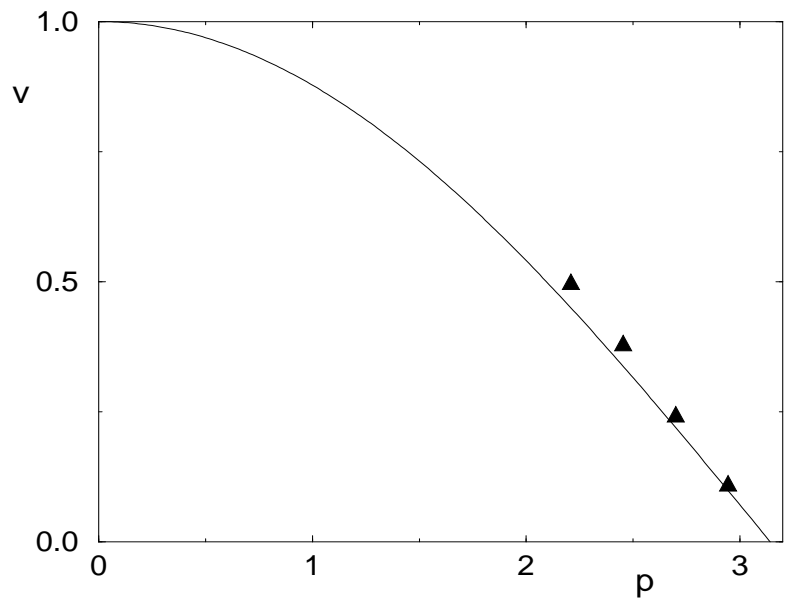}} 
\caption{Velocity of travelling multibreathers versus the wavenumber of the 
forcing. The solid line is the group velocity of linear waves.}
\label{velocity}
%\bigskip
\end{figure}

In Fig.~\ref{velocity} we plot the speed of the travelling multibreather
as a function of the wavenumber of the forcing $p$-mode, which compares well with
the group velocity of the corresponding linear waves, showing that nonlinear
effects are negligible in this parameter range.

\section{Conclusions}
\label{Sec4}
We have shown how to extend the analysis of the $\pi$-mode solution
in an ``externally" forced and damped FPU 1D lattice performed in
Ref.~\cite{ssr} to that of
a generic $p$-mode solution with $\pi/2 < |p| < \pi$. Since the carrier
wave has a nonzero velocity, we find, for weak forcing, solutions corresponding
to {\it nonlinear travelling waves}. These solutions become unstable
above a critical forcing $f_{cr}$, of which an analytical 
expression can be derived in terms of the parameters of the system:
the  frequency of the forcing $\omega$, the damping constant $\gamma$
and the wavenumber $p$.
Above $f_{cr}$ the $p$-mode solution is modulationally unstable and the
system generates complex patterns after nonlinear saturation:
{\it travelling modulated waves} for in-band frequency forcing $\omega <\omega_*$
or {\it travelling multibreathers} for out-band forcing $\omega > \omega_*$.
We suggest that the multibreather pattern could be described by the
soliton solutions of a driven-damped nonlinear Schr\"odinger
equation.
Travelling soliton trains have been also recently observed for the
weakly damped parametrically driven NLS equation~\cite{martel}.
Our study also hints at the existence of similar mechanisms 
for the generation of moving breathers
in more complex physical systems, which are
known to be governed by the same driven-damped NLS or Sine-Gordon equation, e.g. 
one dimensional easy-axis magnetic structures~\cite{lai}.

\begin{ack}
This work is supported by the EC network LOCNET, Contract No. HPRN-CT-1999-00163
and by the MURST-COFIN00 project {\it Chaos and localization in classical and quantum
mechanics}. S.L. acknowledges useful discussions with C. Martel as well as
partial financial support from the Region Rh\^one-Alpes, France.
\end{ack}

\newpage

\end{document}